\documentclass[aps,prb,twocolumn]{revtex4}
\usepackage{graphicx}
\usepackage{amsmath}
\usepackage{bm}% bold math
\def\paraH2{{\it p}-H$_2$}
\def\he4{He$^4$}
\def\Am2{\AA$^{-2}$}
\def\gapx{\lower 2pt \hbox{$\buildrel>\over{\scriptstyle{\sim}}$}}
\def\lapx{\lower 2pt \hbox{$\buildrel<\over{\scriptstyle{\sim}}$}}

\begin{document}

\title{Absence of superfluidity in a parahydrogen film \\  intercalated within a crystal of Na atoms} 

\author{Massimo Boninsegni}
\affiliation{Department of Physics, University of Alberta, Edmonton, Alberta, Canada T6G 2J1}

\date{\today}

\begin{abstract}
A recent claim of possible superfluid behaviour of parahydrogen films intercalated within a crystalline 
matrix of Na atoms is examined. 
Quantum Monte Carlo simulations at finite temperature yield strong numerical evidence that the system forms at
low temperature a {\em non-superfluid} crystalline phase, commensurate with the underlying impurity lattice. The physics of this system is therefore qualitatively identical to that observed in similar settings, extensively studied in precedence.  Comparison of 
numerical results obtained here, with those of the reference in which the prediction of superfluidity (disproven here) was made, 
points to likely bias in the computational methodology adopted therein.
\end{abstract}
\maketitle

\section{Introduction}

Condensed parahydrogen (\paraH2) was predicted over forty years ago to undergo a superfluid (SF) transition at low temperature 
($T \lesssim 6$ K). The physical argument is very simple, and consists of treating the system as a non-interacting ensemble of  \paraH2 molecules,  regarded as point-like Bose particles of spin zero.\cite{ginzburg72} Such a relatively 
crude approach  provides a reasonably accurate estimate of the superfluid transition temperature $T_c$ of 
liquid \he4 at saturated vapour pressure;  the reason is that the equilibrium phase of \he4 is a liquid in the $T\to 0$ limit, 
and retains the most important qualities of a non-interacting Bose gas, notably it undergoes  Bose-Einstein Condensation.\cite{feynman}
\\ \indent
However, bulk \paraH2 crystallizes at low $T$, in spite of the low mass of the molecules, due to the depth of the attractive well of the intermolecular potential, roughly three times that between two helium atoms. Indeed, the low temperature equilibrium phase of \paraH2 is theoretically predicted to be a (non-superfluid) crystal  in reduced dimensions as well, with not 
even a {\it metastable} fluid phase.\cite{boninsegni04,boninsegni13} 
There is fairly robust numerical evidence\cite{sindzingre,fabio,fabio2}  of superfluidity in small \paraH2 clusters (thirty molecules or less), 
which remain ``liquidlike" at low $T$, leading to the belief that a bulk superfluid phase should be observable, if crystallization of the fluid phase could be suppressed. However, this goal has been so far achieved only for droplets of up to approximately $10^4$ molecules;\cite{vilesov} 
 none of many experimental attempts to stabilize a bulk liquid phase\cite{bretz81,maris86,maris87,schindler96} has so far met with success.
\\ \indent
The  suggestion was made, almost two decades ago,\cite{gordillo97}  that SF might occur in a (quasi) 
two-dimensional (2D) \paraH2 fluid embedded in a crystalline matrix of 
Alkali atoms. The contention is that  the presence of  the underlying lattice of foreign atoms,
incommensurate with the equilibrium crystal structure of pure \paraH2,   could possibly 
cause a substantial reduction of the equilibrium density of the 2D fluid of 
\paraH2 molecules, stabilizing a  liquid phase. \\ Path Integral Monte Carlo (PIMC) 
simulations appeared to support such a scenario, providing evidence of  a superfluid transition at $T$ $\sim$ 1 K. Subsequent studies,\cite{njp,turnbull} however, disproved such a conclusion, showing it to be merely an artifact of simulations carried out on systems of extremely small size 
($\sim 10$ particles). In actuality, the equilibrium phase is a non-superfluid crystal, commensurate with the underlying impurity lattice, with a 10/3 density ratio; its unambiguous observation by computer simulation requires that the simulated system comprise a sufficient ($\sim$ 120) number of \paraH2 molecules.\cite{turnbull} A broader conclusion of those studies was that, although it is true that confinement and disorder can indeed lead to novel phases of matter\cite{stan}  (and indeed the superfluid response of \paraH2 clusters can be enhanced in confinement\cite{omiyinka}), the strong propensity of \paraH2 to solidify renders it  exceedingly unlikely that one may arrive at a SF phase in this way, as a commensurate crystal is the only additional phase that can result from the presence of an external periodic potential.
\\ \indent
Recently, however, the claim of a possible superfluid phase of \paraH2 in the same physical setting was reiterated.\cite{cazzorla} Specifically, it was contended that  ``fine tuning"  the potential  describing the interaction of \paraH2 molecules with the impurities (specifically, choosing its parameters to correspond roughly to the interaction of a \paraH2 molecule with a Na atom) has the effect of enhancing ``fluidlike'' behaviour of the system, leading to a nonzero superfluid response at $T=0$. This conclusion based on (ground state) Diffusion Monte Carlo (DMC) simulations of the same model system studied in Refs. \onlinecite{gordillo97,njp,turnbull}, only with a different choice of potential parameters.
\\ \indent
In order to provide an independent check of this surprising and counterintuitive prediction, we have carried out Quantum Monte Carlo simulations of the same system studied in Ref. \onlinecite{cazzorla}, using the same potentials utilized therein. 
We made use of a different computational methodology, namely we used the continuous-space Worm Algorithm. This (Monte Carlo) technique  provides accurate estimates of thermodynamic properties of Bose systems at finite temperature, and has the distinct advantage of not relying on any {\it a priori} input, such as a trial wave function in the case of DMC. We carried out simulations down to a temperature $T$=0.125 K, which, as we argue below, is low enough to regard results as representative of ground state physics.
\\ \indent
Our results are in disagreement with the predictions of Ref. \onlinecite{cazzorla}. We show that the physical behavior of this system is qualitatively identical with that observed in all previous studies  with different potentials,\cite{njp,turnbull} i.e., the only different phase that forms, with respect to purely 2D \paraH2, is the  
non-superfluid crystalline phase described above, commensurate with the underlying impurity lattice.
This can be established both by an examination of the energetics, as well as from the direct computation of cogent quantities such as the pair correlation function, the superfluid density, the Lindemann ratio, as well as the frequency with which exchanges of indistinguishable particles occur, and their physical character. \\ \indent
In other words, no evidence is seen of  the change in the physics of the system proposed in Ref. \onlinecite{cazzorla},  allegedly arising from a weaker interaction   between a  \paraH2 molecule and the impurity atom than that considered in previous studies. 
On the contrary, as already suggested in Ref. \onlinecite {njp}
the use of different parameters to characterize the interaction between \paraH2 molecules and impurity atoms brings about no new physics whatsoever, essentially because no different physics is possible in the setup considered here.\cite{dang}
We argue that the incorrect prediction of superfluid behaviour made in Ref. \onlinecite{cazzorla} is in part a consequence of the failure to identify the equilibrium phase, but also of inherent bias of DMC. 
\\ \indent
The remainder of this paper is organized as follows: in Sec. \ref{mc} we introduce the model and provide computational details; in   Sec. \ref{sere} we illustrate our results and provide a theoretical interpretation. Finally, we outline our conclusions in Sec. \ref{conc}.

\section{Model and Calculation}\label {mc}
As mentioned in the Introduction, we model our system of interest as in all previous comparable studies, namely Refs. \onlinecite{gordillo97,njp,turnbull,cazzorla}. We considered a collection of $N$ point-like particles (\paraH2 molecules) of mass $m$, moving in two dimensions in the presence of an external potential arising from a lattice of static, identical impurities. The system is enclosed in a rectangular simulation cell of sides $L_x = 60$ \AA, $L_y=51.9615$ \AA\ (and area ${\cal A}=L_x\times L_y$), with periodic boundary conditions in all directions.  The nominal 2D  density (coverage) of \paraH2 is $\theta=N/\cal A$.
The quantum-mechanical Hamiltonian of the system is the following:
\begin{equation}\label{ham}
\hat {\cal H} = -\frac{\hbar^2}{2m}\sum_{i=1}^N\nabla_i^2 +\sum_{i<j}V(r_{ij})+\sum_{i\sigma}U(|{\bf r}_i-{\bf R}_\sigma|)
\end{equation}
Here, $V$ is the interaction potential between any two \paraH2 molecules, only depending on their relative distance $r_{ij}\equiv |{\bf r}_i-{\bf r}_j|$; 
the accepted Silvera-Goldman\cite{SG} potential is used to describe this interactions. 
The system also includes $M$ impurities, positioned at regular lattice sites ${\bf R}_\sigma$, with $\sigma=1,2,...,M$ of a triangular lattice, with lattice constant 10 \AA. $M=36$ in this study, i.e., the density of impurities is  $M/{\cal A} = 0.01155$ \AA$^{-2}$.
\\ \indent
The  interaction between a \paraH2 molecule and an impurity (i.e., the $U$ term in (\ref{ham})) is described by a Lennard-Jones potential with parameters $\epsilon=30$ K and $\sigma$=4.14 \AA, i.e., as suggested in Ref. \onlinecite{cazzorla}, where it is claimed to provide a reasonably realistic description of the interaction of a \paraH2 molecule with a Na atom.  It is worth noting that, as suggested in Ref. \onlinecite{gordillo97}, it may be feasible to produce a lattice of Alkali atoms such as the one described here, by adsorbing fractions of a monolayer of Alkali metal atoms (Rb, Cs, and K) onto a Ag(111) or on a graphite substrate.\cite{diehl,diehl2}
\\ \indent
We studied the low temperature physical properties of the system described by Eqs. (\ref {ham}) by means of first principle computer simulations based on the Worm
Algorithm in the continuous-space path integral representation.\cite{worm,worm2} Because this well-established computational methodology is thoroughly described elsewhere, we do not review it here. The most important aspects to be emphasized here, are that it enables one to compute thermodynamic properties of Bose systems at finite temperature, directly from the microscopic Hamiltonian, in particular  energetic, structural and superfluid properties, in practice with no approximation. Technical details of the simulation are standard, and we refer the interested reader to Ref. \onlinecite{worm2}. We used the standard high-temperature approximation for the many-particle propagator accurate up to order $\tau^4$, and all of the results reported here are extrapolated to the $\tau\to 0$ limit;  in general, we found that a value of the imaginary time step 
$\tau=1/320$ K$^{-1}$ yields estimates that are indistinguishable from the extrapolated ones, within the statistical errors of the calculation. We obtained results in the temperature range $0.125$ K $\le T \le 1 $ K. 
\section{Results}\label{sere}
\begin{figure}[h]
\centerline{\includegraphics[height=2.4in]{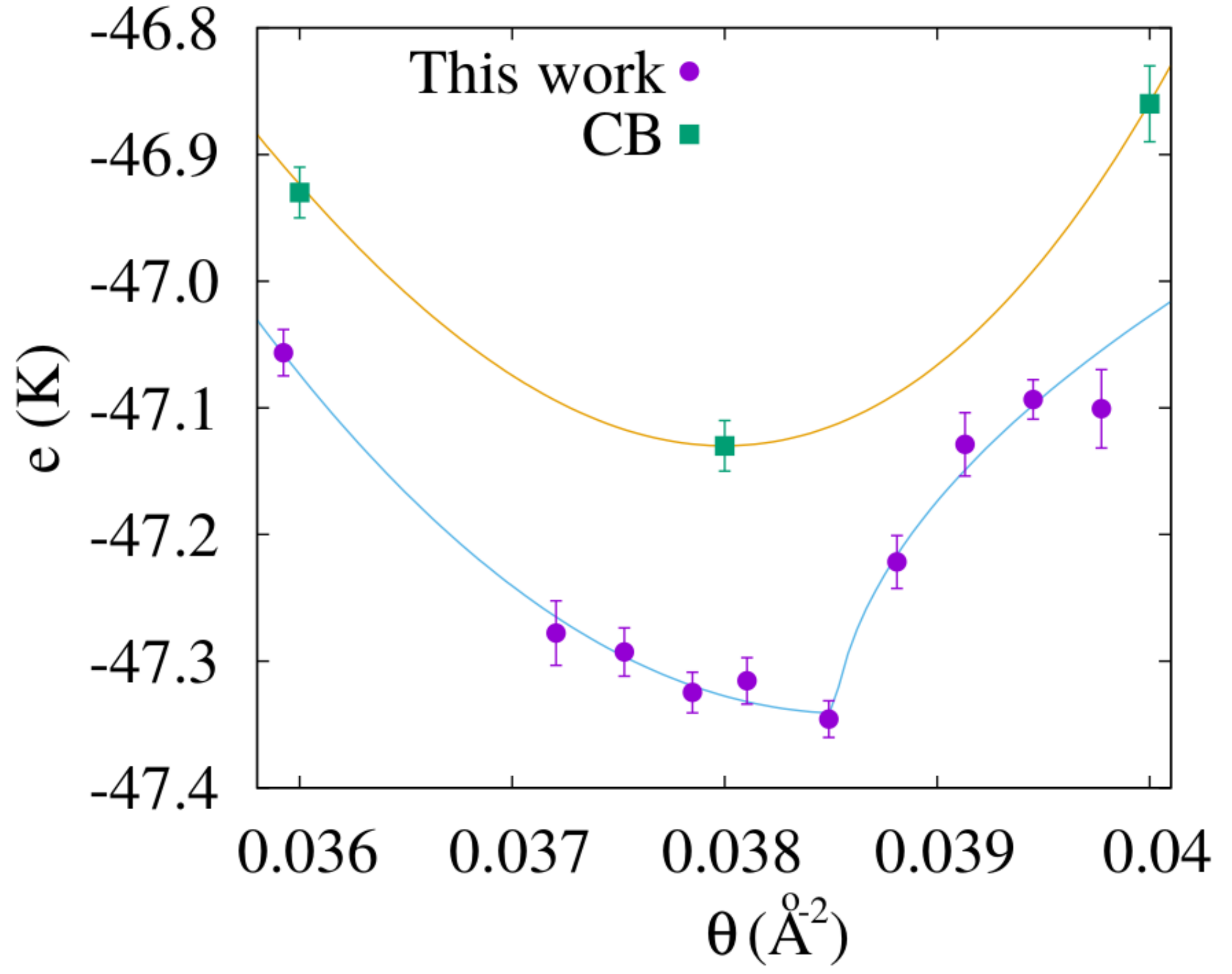}}
\caption{Energy per \paraH2 molecule (in K) at different 2D densities (in \Am2). The concentration of impurities is 0.01155 \Am2. Circles: this work, $T$=1 K, 36 impurities. Boxes: DMC results of Ref. \onlinecite{cazzorla}, quoted therein as ground state estimates, 30 impurities. Solid line going through the boxes is the fit to the DMC data proposed in Ref. \onlinecite{cazzorla}, whereas that through the 
circles was obtained in this work. Separate curves were obtained above and below the equilibrium coverage $\theta_0=0.0385$ \Am2. }
\label{f1}
\end{figure}
Fig. \ref{f1} shows computed energetics of 2D \paraH2 films at different 2D
coverages $\theta$. Our results are for a temperature $T$=1 K; we find that the results for all  relevant physical quantities do not change significantly in the temperature range considered here. The contribution to the potential energy from particles outside the main simulation 
cell can be estimated at less than half of our typical statistical errors (or the order of 0.02 K). The first remark is that the energy per molecule $e(\theta)$ displays a well-defined minimum at $\theta_0=0.0385$ \Am2, with an energy per molecule $e=-47.353(16)$ K.
For our simulated system, the coverage $\theta_0$ corresponds to   a  density equal to {\it precisely} 10/3 that of the underlying impurity lattice. \\ \indent
As mentioned in the Introduction, this is exactly what already observed in previous studies of this system, albeit with a different choice of parameters of the interaction $U$, and indeed the $e(\theta)$ curve shown in Fig. \ref{f1} is basically identical to that of the inset of 
Fig. 1 of Ref. \onlinecite{turnbull}. In particular, its abrupt\cite{crespi} change of slope at $\theta_0$ is not only consistent with the equilibrium phase being  { commensurate}, but also suggests that interstitial doping  will not result in a homogeneous phase, but rather in the coexistence of two commensurate phases. As expected and shown below, the equilibrium phase at coverage $\theta_0$ displays the same crystalline arrangement of \paraH2 molecules already seen on different substrates with the same geometry but different parameters of the interaction term $U$.
\\ \indent
\begin{figure}[th]
\centerline{\includegraphics[height=2.5 in]{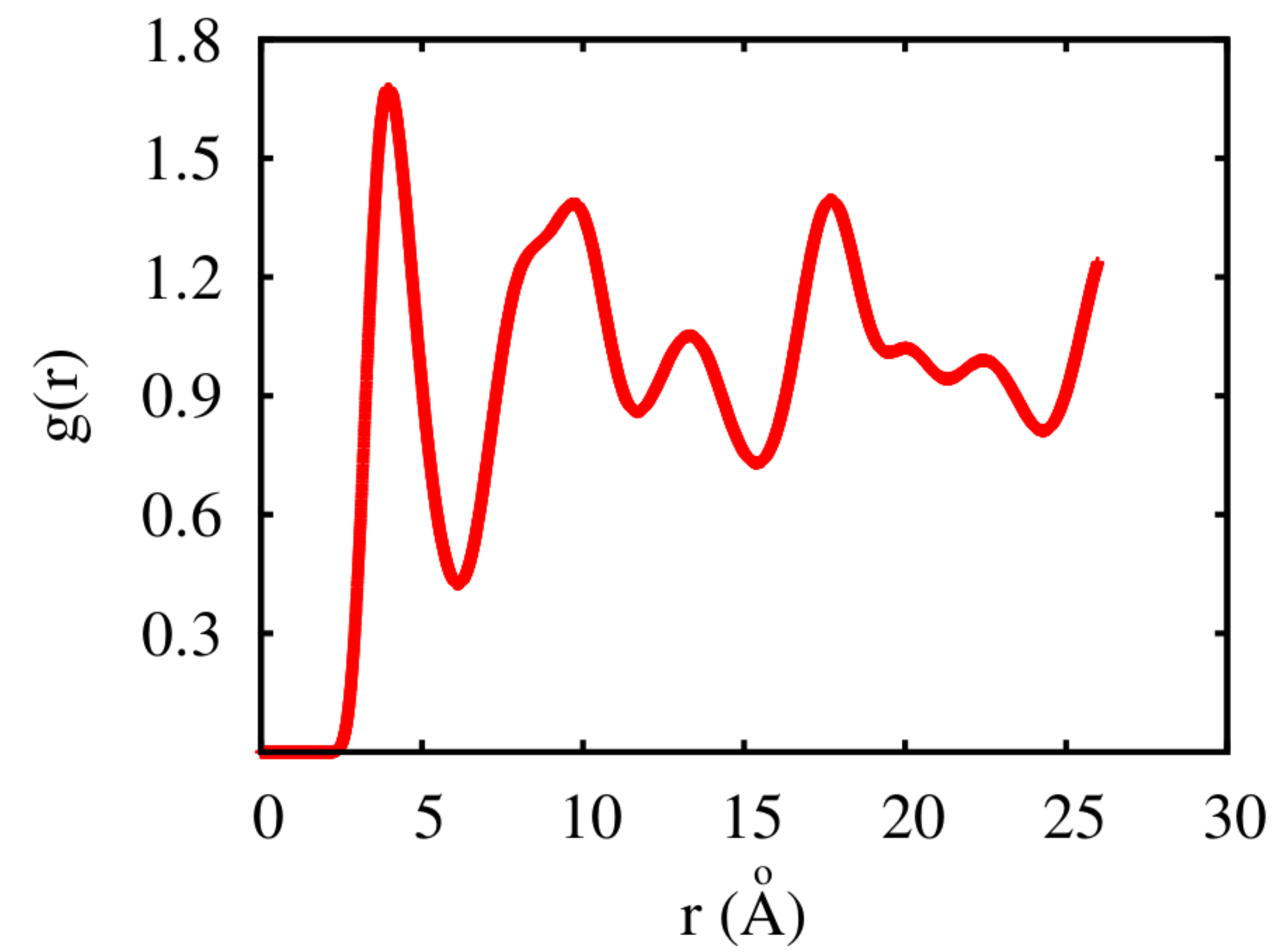}}
\vspace*{8pt}
\caption{Pair correlation function of \paraH2 molecules, computed  at $T$=0.125 K and at the equilibrium density ($\theta_0$= 0.0385 \Am2) on a 120-particle system. Distances ($r$) are given in \AA.}
\label{f22}
\end{figure}
\begin{figure}[th]
\centerline{\includegraphics[height=2.4 in]{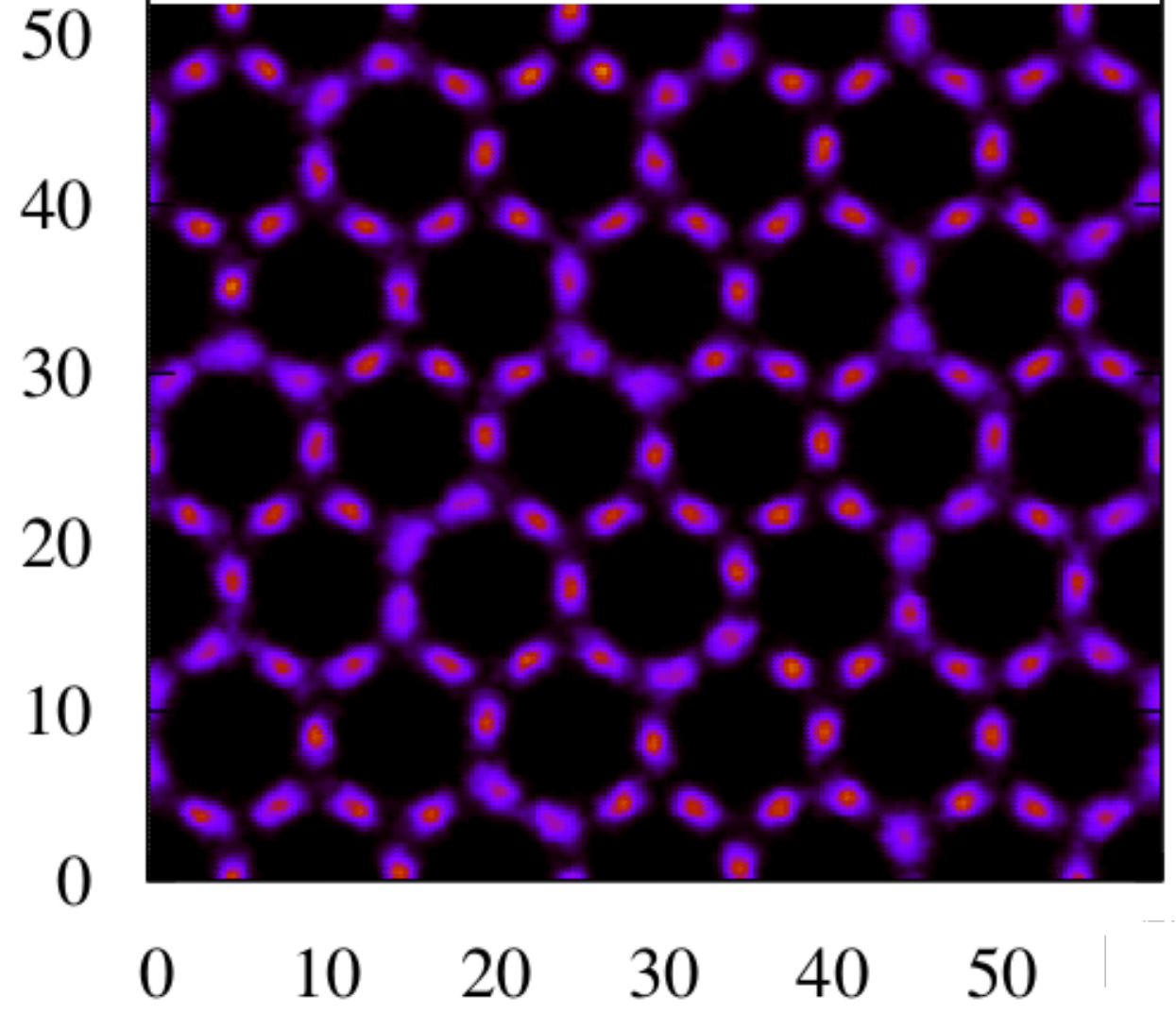}}
\caption{{\it Color online.} Density snapshot of the system    at the equilibrium density $\theta_0$, at $T$=0.125 K.  All lengths are in \AA. Impurity atoms are not shown for clarity, there is one in the middle of each ring
of \paraH2 molecules. }
\label{f2}
\end{figure}
Comparison of our results with those of Ref. \onlinecite{cazzorla} (henceforth referred to as CB) shows obvious, significant qualitative and 
quantitative differences between the two studies. First, the energy values obtained in this work at $T$=1 K are consistently {\em lower} than the (supposedly ``exact") DMC ground state estimates, by approximately 0.2 K. For example, the value of the energy per \paraH2 molecule found here at $\theta_1=0.038$ \Am2 is $-47.31(1)$ K, as opposed to  $-47.13(2)$ K reported in Ref. \onlinecite{cazzorla} for the same coverage. \cite{cabuzzo}
Second, and more important, there are not enough points in the $e(\theta)$ curve of CB, not only  to obtain a precise estimate of the equilibrium coverage (quoted in CB at $\theta=0.0381$ \Am2, i.e., away from commensuration), but also to capture important details of the shape of the curve, as seen in Fig. \ref{f1}. Thus, the fit to the DMC data obtained in CB is  misleading.
\\ \indent
Fig. \ref{f22} shows the pair correlation function $g(r)$ for the \paraH2 molecules at the equilibrium coverage $\theta_0$, at the lowest
temperature considered here, namely $T=0.125$ K. Our result is, again, virtually identical to that of Ref. \onlinecite{turnbull}, and also reasonably close to that given in CB at the slightly lower coverage
$\theta_1$,  featuring considerable structure, as expected from the presence of the impurity lattice. In principle, of course, there is no reason why the adsorbed film may not display some of the qualities of a liquid, even though its density will inevitably not be uniform but reflect the underlying external potential arising from the impurity atoms. In particular, molecules may still enjoy a great deal of mobility, and quantum-mechanical exchanges of indistinguishable particles, which underlie superfluidity, may still occur.
This is, however, not the case, as we now show.
\\ \indent
Fig. \ref{f2} shows a typical 
instantaneous density snapshot of the  system at the equilibrium density $\theta_0$, at a temperature $T$=0.125 K. It can be regarded as representative of the physics of the system at  $T \le 1$ K, as it is qualitatively identical to many other similar snapshots, collected at random times in the  course of long simulations at different temperatures. 
Aside from the arrangement of \paraH2 molecules on a regular  {\it kagom\'e}  lattice, which  is clear, it is worth noting that molecules are very nearly ``pinned" at lattice sites, with little or no overlap  between the delocalization clouds of adjacent molecules, i.e., that quantum-mechanical exchanges are all but suppressed. Indeed, the only (very infrequent) permutation of indistinguishable molecules that is observed in the simulations, in the temperature range explored here, is simply a rotation of the seven molecules on one of the rings of the lattice. It is important to stress that molecules are {\it not}  placed as shown in Fig. \ref{f2} at the start of the simulation, but rather such an arrangement appears spontaneously, even if molecules are initially positioned differently (e.g., on a uniform triangular lattice).\\ \indent
The crystalline, insulating nature of this system can be quantitatively, conclusively established through the so-called {\em Lindemann ratio}, namely the ratio of the rms excursion $u$ of molecules away from their equilibrium points and the mean intermolecular distance. This quantity can easily be computed with the methodology utilized here, and its value at $T$=0.125 K is $\sim$ 0.28; for comparison, in the 2D bulk crystalline phase of parahydrogen at its equilibrium density,\cite{boninsegni04} namely 0.067 \Am2 at $T$=0.5 K is $\sim$ 0.33. Thus, the results of this study yield strong evidence that the system forms a commensurate crystal, with no evidence of liquidlike behavior.

\begin{figure}[th]
\centerline{\includegraphics[height=2.4 in]{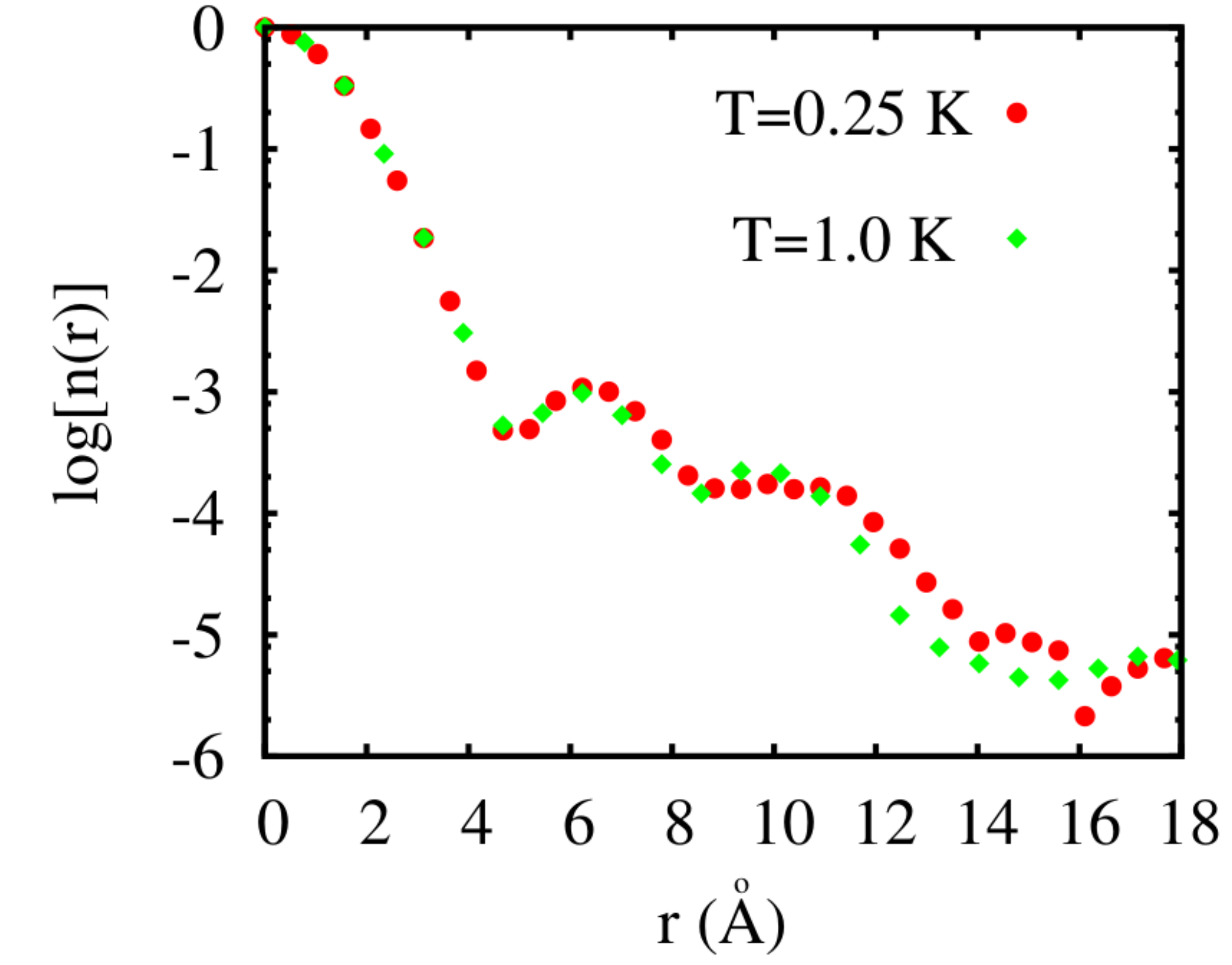}}
\caption{{\it Color online.} One-body density matrix (log scale, base 10) for the system at the equilibrium density $\theta_0$, computed at the two temperatures $T$=1 K (diamonds) and $T$=0.25 K (circles). Statistical errors are of the order of the symbol sizes.}
\label{obdm}
\end{figure}

It is in principle not impossible for a system to feature at the same time superfluid and crystalline properties. However, in this case the strong suppression of quantum-mechanical exchanges, and in particular the absence of long permutations spanning the whole system, results in a  value of zero of the superfluid density, down to the lowest temperature considered here; concurrently, and expectedly, 
the one-body density matrix displays a temperature-independent exponential decay with distance, as shown in Fig. \ref{obdm}. In other words, nothing points to a qualitative change of physics in the limit in which $T\to 0$.
This conclusion is also consistent with the repeated observation that for a system of 
hard-core bosons in the presence of a periodic external potential, the superfluid fraction vanishes at commensurate densities.\cite{dang}
\\ \indent
This conclusion is manifestly at variance with the claim made in CB that the ground state of the system is a {\em superfluid liquid}, with a value of the superfluid fraction approaching 30\%. The most obvious explanation for the different physical behavior observed here, is simply  that  they missed the commensurate crystalline phase. While superfluidity is absent at the commensurate, equilibrium coverage $\theta_0$, the system might be in a metastable superfluid phase below the equilibrium density, and indeed this was the claim made in the original study by Gordillo and Ceperley.\cite{gordillo97,zuppo}  However, much like in Ref. \onlinecite{turnbull} in which a different set of parameters for the potential $U$ in  Eq. (\ref{ham}) was utilized, calculations carried out in this work  at {\it all} coverages, including  below $\theta_0$ consistently yielded no evidence of anything resembling a ``liquid" phase;\cite{preci} on the contrary,  the same non-superfluid, commensurate 
crystal illustrated above for $\theta=\theta_0$ was observed,  with no superfluid signal down to $T=0.125$ K. Indeed, there is no evidence at all that the different choice of parameters for $U$ leads to grater mobility of the \paraH2 molecules, as proposed in CB.
Since both calculations make the claim of being numerically ``exact" (meaning, errors are only statistical in nature and can thus be rendered arbitrarily small by employing a sufficient amount of CPU time), any numerical discrepancy or physical  should be carefully
examined and resolved. 
\\ \indent
The first thing to mention is that, although it is often advertised as exact, the DMC method is in fact  affected by a bias associated to the trial wave function out of which the ground state is projected, as well as by the necessarily finite population of random walkers  utilized. 
Such a bias, often very difficult to remove (even with very long computer runs\cite{moroni})  has led to several  DMC predictions of liquidlike behavior or superfluidity of various systems that were eventually disproven.\cite{boninsegni13,jltp,happacher}  In general, overwhelming evidence now suggests that finite temperature methods constitute a superior option to investigate Bose systems -- even their ground state. \\ \indent
The comparison of energy estimates of Fig. \ref{f1},  showing finite temperature results consistently, significantly below DMC ones,  suggests  that the prediction of superfluid behavior
made in CB is merely an artifact of the DMC methodology utilized therein, specifically of the failure of the DMC projection to converge
to the true ground state in the relevant range of coverage (phrased alternatively, failure to remove entirely the bias associated to the trial wave function).
\\ \indent
One could argue that  the superfluid transition predicted by CB may simply occur at lower temperatures than the ones considered in this work. However, a hypothetical superfluid transition should still conform to the accepted Kosterlitz-Thouless paradigm, with the well-known universal jump condition.\cite{KT} On assuming a value of superfluid fraction at the transition temperature $T_c$ equal to one half of the saturation value (0.3) claimed in CB, one comes up with $T_c\sim 0.2$ K, i.e., barring some exceedingly unlikely scenario of melting of the communsurate crystal at very low temperature, evidence of it should definitely be seen in our study. In particular, the one-body density matrix should display a marked dependence on temperature, which is not seen here.

\section{CONCLUSIONS}\label{conc}
Based on an extensive computational study of two-dimensional {\it para}-Hydrogen embedded in a 
crystalline matrix of Na atoms, modeled in exactly the same way as in
a previous study, we conclude that this system is {\it not } a candidate for
observing superfluidity in \paraH2. At low temperature, the system forms 
instead a 2D crystal, commensurate with the underlying lattice of impurities. 
We have presented results for triangular impurity lattices, but the same 
results were seen with other lattices as well (e.g., rectangular). 
In striking disaccord with what proposed in Ref. \onlinecite{cazzorla}, no qualititative nor quantitative change is brought about by tweaking  the parameters of the potentials  used to describe the interaction of a \paraH2 molecule with an impurity atom. As established in all previous studies,\cite{njp,turnbull} when prevented by an impurity lattice from forming their preferred crystalline arrangement,  \paraH2 molecules simply do the ``next best thing", namely form a crystal {\it commensurate} with such an underlying lattice. 
The 
finite superfluid signal obtained by other authors can be attributed to the  bias inherent in the computational methodology adopted in Ref. \onlinecite{cazzorla}.
More generally, this study confirms the physical conclusion of absence of superfluidity in 2D \paraH2 in the presence of external periodic potentials.

\section*{Acknowledgements}

This work was supported  by the Natural Science
and Engineering Research Council of Canada. Computing support of Westgrid is gratefully acknowledged.


\begin{thebibliography}{0}
\bibitem{ginzburg72}
V. L. Ginzburg and A. A. Sobyanin,  JETP Letters {\bf 15}, 242 (1972).
\bibitem{feynman}
R. P. Feynman,  Phys. Rev. {\bf 91}, 1291 (1953).
\bibitem{boninsegni04} M. Boninsegni, Phys. Rev. B {\bf 70}, 125405 (2004).
\bibitem{boninsegni13}
M. Boninsegni, Phys. Rev. Lett. {\bf 111}, 235303 (2013).
\bibitem{sindzingre} 
P. Sindzingre, D. M. Ceperley and M. L. Klein, 
Phys. Rev. Lett. {\bf 67}, 1871 (1991).
\bibitem{fabio}
F. Mezzacapo and M. Boninsegni, Phys. Rev. Lett. {\bf 97}, 045301 (2006).
\bibitem {fabio2}
F. Mezzacapo and M. Boninsegni, Phys. Rev. A {\bf 75}, 033201 (2007).
\bibitem{vilesov}
K. Kuyanov-Prozument and A. F. Vilesov, Phys. Rev. Lett. {\bf 101}, 205301 (2008).
\bibitem{bretz81}
M. Bretz and A. L. Thomson,
Phys. Rev. B {\bf 24}, 467 (1981).

\bibitem{maris86}
G. M. Seidel, H. J. Maris, F. I. B. Williams and J. G. Cardon,
Phys. Rev. Lett. {\bf 56}, 2380 (1986).

\bibitem{maris87}
H. J. Maris, G. M. Seidel and F. I. B. Williams, 
Phys. Rev. B {\bf 36}, 6799 
(1987).

\bibitem{schindler96}
M. Schindler, A. Dertinger, Y. Kondo and F.  Pobell, 
Phys. Rev. B {\bf 53}, 11451 (1996).

\bibitem{gordillo97}
M. C. Gordillo and D. M. Ceperley, Phys. Rev. Lett.  {\bf 79}, 3010 (1997).
\bibitem{njp}
M. Boninsegni, New J. Phys. {\bf 7}, 78 (2005).
\bibitem{turnbull}
J. Turnbull and M. Boninsegni, Phys. Rev. B {\bf 78}, 144509 (2008).
\bibitem{stan}
See, for instance, G. Stan, S. Gatica, M. Boninsegni, S. Curtarolo and M. W. Cole, Am. J. Phys. {\bf 67}, 1170 (1999), and references therein.
\bibitem{omiyinka}
T. Omiyinka and M. Boninsegni, Phys. Rev. B {\bf 90}, 064511 (2014).
\bibitem{cazzorla}
C. Cazorla and J. Boronat, Phys. Rev. B {\bf 88}, 224501 (2013).
\bibitem{dang}
L. Dang and M. Boninsegni, Phys. Rev. B {\bf 81}, 224502 (2010).
\bibitem{SG}
I. Silvera and V. Goldman, J. Chem. Phys. {\bf 69}, 4209 (1978).
\bibitem{diehl}
J. Cui, J. D. White, R. D. Diehl, J. F. Annett and M. W. Cole,  Surf. Sci. {\bf 279}, 149 (1992).
\bibitem{diehl2} 
G. S. Leatherman and R. D. Diehl, Phys. Rev. B {\bf 53}, 4939 (1996).

\bibitem{worm}
M. Boninsegni, N. Prokof'ev and B. Svistunov, Phys. Rev. Lett. {\bf 96}, 070601 (2006).
\bibitem{worm2}
M. Boninsegni, N. Prokof'ev and B. Svistunov, Phys. Rev. E {\bf 74}, 036701 (2006).
\bibitem{crespi}
M. Boninsegni, S.-Y. Lee and V. H. Crespi, Phys. Rev. Lett.  {\bf 86}, 3360 (2001).
\bibitem{cabuzzo}
{Although an overall shift may not, {\em per se},  lead to a different physical picture, this discrepancy is noteworthy in that  {\it a}) it  is not small (as seen on the scale of Fig. \ref{f1}),  and {\it b}) it represents yet another instance of  finite temperature methods yielding more accurate {\it ground state} estimates than DMC. See, for instance, Refs. \onlinecite{boninsegni13} and \onlinecite{moroni}.}

\bibitem{moroni}
M. Boninsegni and S. Moroni, Phys. Rev. E {\bf 86}, 056712 (2012).
\bibitem{zuppo}
The large superfluid response reported by CB cannot arise from doping a commensurate crystal  with a mere $\sim 1$\% vacancy.
\bibitem{preci}
The adjective ``liquid" used in Ref. \onlinecite {cazzorla} seems arbitrary, as  the trial wave function utilized in the projection breaks translational invariance, in a way that reflects the underlying impurity lattice, even though molecules are not  pinned at Na  sites.

\bibitem{jltp}
M. Boninsegni, J. Low Temp. Phys. {\bf 165}, 67 (2011).
\bibitem{happacher}
J. Happacher, P. Corboz, M. Boninsegni and L. Pollet, Phys. Rev. B {\bf 87}, 094514 (2013).
\bibitem{KT} J. M. Kosterlitz and D. J. Thouless, J. Phys. C {\bf 6}, 1181 (1973).
\end{thebibliography}
\end{document}